# The relativistic Pauli equation

D. H. Delphenich[†]
Kettering, OH USA 45440

**Abstract.** After discussing the way that $\mathbb{C}^2$ and the algebra of complex 2×2 matrices can be used for the representation of both non-relativistic rotations and Lorentz transformations, we show that Dirac bispinors can be more advantageously represented as 2×2 complex matrices. One can then give the Dirac equation a form for such matrix-valued wave functions that no longer necessitates the introduction of gamma matrices or a choice for their representation. The minimally-coupled Dirac equation for a charged spinning particle in an external electromagnetic field then implies a second order equation in the matrix-valued wave functions that is of Klein-Gordon type and represents the relativistic analogue of the Pauli equation. We conclude by presenting the Lagrangian form for the relativistic Pauli equation.

Contents

Page

**1. Introduction.** One of the first challenges to the success of the Schrödinger equation in atomic physics was the splitting of the lines in the spectra of atomic photons in the presence of magnetic fields, namely, the Zeeman effect. The subsequent discovery by Stern and Gerlach of a non-zero magnetic dipole moment for the electron that seemed to exist in two distinct states then spawned a lot of theoretical research into how one would then have to modify the mathematical models for the electron in order to account for that fact.

---

[†] E-mail: david_delphenich@yahoo.com, Website: neo-classical-physics.info

Frenkel [**1**] devised a relativistic classical model for the motion of a spinning electron in which the Lorentz force imparted on it by an external electromagnetic field would then be augmented by a torque on it that was due to the coupling of the magnetic moment to the external magnetic field. More generally, one could image a coupling of the electromagnetic moment to an external electromagnetic field. However, the fact that the electron seemed to have no measurable electric dipole moment suggested to Frenkel that one might impose the constraint that the electromagnetic moment was, in fact, purely magnetic, which is then usually referred to as the "Frenkel constraint."

As far as quantum physics was concerned, it was the Uhlenbeck-Goudsmit hypothesis [**2**] that defined the first major advance in the theory. Their hypothesis was that the electron's magnetic moment might be due to it having an intrinsic angular momentum – or *spin* – that existed only in two states (which are usually called "up" and "down") and the magnetic moment was proportional to that spin by way of a factor that is called the *Bohr magneton*. It is important to observe that although the original conception of intrinsic angular momentum was often kinematical in nature, it eventually became clear that one should not think of spin as an actual rotational motion of the electron, so much as an artifact of the field space in which the wave function of the electron takes it values; i.e., it relates to the *weight* of the representation of the rotation or Lorentz group in the field space.

The notion of introducing non-trivial representations of the rotation or Lorentz group into the quantum wave function of the electron was largely due to Wolfgang Pauli [**3**], who first replaced the one-dimensional complex vector space $\mathbb{C}$ of the usual Schrödinger wave functions with the two-dimensional space $\mathbb{C}^2$. Since Pauli was still considering the non-relativistic motion of electrons, he only required that $\mathbb{C}^2$ carry a representation of the group $SO(3; \mathbb{R})$ of real, orientation-preserving, three-dimensional Euclidian rotations that comes about by way of the group $SU(2)$ of unitary 2×2 complex matrices of unit determinant that acts on $\mathbb{C}^2$ by matrix multiplication. The fact that the homomorphism $SU(2) \to SO(3; \mathbb{R})$, which amounts to the projection of $S^3$ onto $\mathbb{R}P^3$ topologically, is two-to-one is at the root of the fact that the spin states exist in two discrete levels. Hence, since that is an experimental fact, one way of characterizing the difference between classical and quantum mechanics is that the covering group $SU(2)$ is more physically fundamental than the more geometrically familiar group $SO(3; \mathbb{R})$.

The equation that Pauli came up with for the non-relativistic motion of the electron in the presence of an external magnetic field amounted to the usual Schrödinger equation for a wave function that took its values in $\mathbb{C}^2$ with an extra term that included the work done by the torque that the magnetic field exerted on the electron's magnetic dipole moment. If it were not for the latter term, the equation would reduce to a pair of uncoupled Schrödinger equations for the components of the wave function, and the spin of the wave function would play no essential role. However, the coupling of the two equations comes about due to the fact that the magnetic dipole moment of the electron was described by an algebraic operator on the wave function that took the form of a 2×2 complex matrix with non-vanishing off-diagonal components. Since the complex vector space $M(2; \mathbb{C})$ of 2×2 complex matrices is four-dimensional, one needs four linearly independent matrices in order to define a basis for it. Here, Pauli introduced a choice of basis that consisted of the identity matrix and three Hermitian matrices $\sigma_i$, $i = 1, 2, 3$ that came to be called the *Pauli matrices*.

The fact that the Pauli equation was manifestly non-relativistic seemed to be a major limitation of the model, and the first successful attempt to devise a wave equation for the spinning electron that was Lorentz-invariant came from Paul Dirac [**4**]. Although Klein [**5**] and Gordon [**6**] had previously discussed how the Schrödinger equation might be given a Lorentz-invariant form, since they were still considering wave functions with values in $\mathbb{C}$ they were ignoring the spin of the wave-particle that was being described.

Also, the wave equation that they arrived at, which amounted to essentially the linear wave equation with a mass term, admitted negative (kinetic) energy solutions that – at the time – seemed unphysical, while implying a conserved current that was associated with the $U(1)$ phase invariance of the equation whose temporal term was not positive-definite. Since the statistical interpretation of the quantum wave function demanded that this term represent a probability density function for the position of the electron (which is then assumed to be point-like), and negative probabilities are absurd, this too was seen to be a drawback to the Klein-Gordon equation as a relativistic wave equation for spinning wave-particles.

Of course, one might also say that the character of the conserved current was a *reductio ad absurdum* for the statistical interpretation of the wave function. Indeed, Schrödinger [**7**] had originally imagined that the modulus-squared of the wave function would represent the electric charge density that it was associated with, which also implies that one is no longer considering point-like matter. Some time after Dirac's theory came out, Pauli and Weisskopf [**8**] revisited the Klein-Gordon equation with the original Schrödinger interpretation of the wave function and concluded that it had many advantages from the standpoint of quantum electrodynamics. However, it still had the drawback that it seemed to only relate to the matter waves that had no spin, although this was later seen to be no drawback in the context of massive scalar mesons.

To return to the Dirac theory, Dirac originally set about the task of formulating a Lorentz-invariant equation – or really, *system* of equations – for the spinning electron that would be first-order, in the hopes that such a system would no longer exhibit the negative-energy solutions. He essentially attempted to find a "square root" of the d'Alembertian operator that would reduce the order of the wave equation from two to one. Of course, in the process of defining the coefficients of the resulting operator, he was forced to imbue them with the structure of a Clifford algebra, namely, the Clifford algebra of real Minkowski space. This had the effect that the smallest extension of the field space $\mathbb{C}$ that would carry a faithful representation of that algebra in its matrix algebra was $\mathbb{C}^4$.

However, since one then defines a system of four first-order partial differential equations for the four components of the wave function, one has not truly reduced the order of the system. Thus, as it turned out, the Dirac equation still admitted negative-energy solutions, which was seen to still be a problem for several years until the positron was discovered. Hence, the only remaining physically definitive objections to the Klein-Gordon equation were how to introduce spin into it and whether one would have to consider some other interpretation of the wave function than the statistical one; for instance, the Schrödinger-Pauli-Weisskopf electrical interpretation seems reasonable.

It was clear from the outset that the Dirac theory of the electron involved a high degree of mathematical artifice that many of the physicists of the era found non-intuitive and lacking in immediate relevance to the rest of what was understood in physics at the time. Consequently, many theoretical studies were carried out in the hopes of finding a more physically-meaningful explanation for the introduction of the methods of Clifford algebra into the modeling of quantum phenomena. One of the recurring themes was the search for choices of representation for the four basic $\gamma$ matrices that enter into the Dirac operator that might put the system of equations into a more physically intuitive form, since there is nothing canonical about any particular choice of representation; any Lorentz transformation of one choice of matrices would produce an equally acceptable alternative representation.

One can always write out the four Dirac component equations explicitly – i.e., with no direct reference to the gamma matrices – and that is exactly what Darwin [9] did for his own studies of the equation. In addition to Dirac's own choice of gamma matrices, two other choices that gained some degree of acceptance were those of Hermann Weyl [10] and Ettore Majorana [11]. One feature of both the Dirac and Weyl representations is that since one can actually faithfully represent the Lorentz group in a smaller complex vector space than $\mathbb{C}^4$ – namely, $\mathbb{C}^2$ – the gamma matrices generally take the form of various (signed) permutations of the blocks in the diagonal representation of the Pauli matrices in the algebra $M(4; \mathbb{C})$ of 4×4 complex matrices. One should also observe that since Minkowski space is four-real-dimensional, its Clifford algebra $Cl(4; \eta_{\mu\nu})$ has real dimension $2^4 = 16$, while the algebra $M(4; \mathbb{C})$ has real dimension 32. Thus, one does not actually require the entire algebra $M(4; \mathbb{C})$ in order represent $Cl(4; \eta_{\mu\nu})$, but only a subalgebra of it. Hence, one can already suspect that there is something gratuitous about the usual gamma matrix representation of $Cl(4; \eta_{\mu\nu})$.

The fact that the gamma matrices can take the form of a "twisted" product representation of the Pauli matrices implies that one can also express the Dirac equation as a pair of equations for a pair of Pauli spinors, rather than a single equation for a Dirac bispinor. This represents an appealing compromise between the unphysical abstraction of the Dirac equation in its usual form and the complexity of the system in its Darwin form, which suggests no obvious algebraic basis for the structure of the equations.

Since the Dirac equation is Lorentz-invariant, one also has that the corresponding equations in terms of Pauli spinors are also Lorentz-invariant, by way of the representation of the proper, orthochronous Lorentz group $SO_+(3, 1)$ in $SL(2; \mathbb{C})$, which then acts upon $\mathbb{C}^2$ naturally by matrix multiplication. This representation is essentially the complexification of the representation of $SO(3; \mathbb{R})$ in $SU(2)$, because $SO_+(3, 1)$ is isomorphic to $SO(3; \mathbb{C})$, while $SL(2; \mathbb{C})$ is, in fact, the complexification of $SU(2)$. Hence, since one has a Lorentz-invariant system of two first-order partial differential equations for two Pauli spinors, one wonders about whether one might have obtained them from some relativistic form of the Pauli equation in a manner that relates to the process of going from the Klein-Gordon equation to the Dirac equation – or vice versa – in a more intuitively appealing way.

One sees that first one must address the way that the coupling of spin and an external electromagnetic field comes about as an addition to the usual Klein-Gordon operator that would then couple the two component equations for the Pauli spinor. In order to do this, one could either start with the non-relativistic Pauli equation and examine the way that it might generalize to a relativistic equation when one replaces the non-relativistic equation for total energy with the relativistic one or start with the Dirac equation and examine the form of the second-order equation that results by applying the Dirac operator to both sides; one can also use the prescription of Feynman and Gell-Mann [**12**]. One finds that for any choice of representation for the gamma matrices one decouples the pair of equations into independent second-order relativistic equations for the individual Pauli spinors. One might then think of these equations as the "relativistic Pauli equations." Thus, it is no longer necessary to introduce gamma matrices or choose a representation for them.

Part of the reason for this is that when $\mathbb{C}^4$ is given the algebra of complex quaternions, one can essentially shift some of the algebraic machinery in the Dirac equation from the coefficients of the differential operator to the components of the wave function, in such a way that the 16-real-dimensional Clifford algebra of real Minkowski space, much less the 32-real-dimensional Clifford algebra of complex Minkowski space becomes an inefficient way of dealing with the essential elements of the wave functions of relativistic spinning particles. However, rather than discuss the algebra of complex quaternions here (one might confer the author's monograph [**13**], which also includes numerous classical references), we shall discuss a compromise measure between the usual physics of Dirac wave functions and the abstraction of algebras, namely, wave functions that take their values in the algebra $M(2; \mathbb{C})$ of complex 2×2 matrices, which is, in fact, isomorphic to the algebra of complex quaternions. As a result, our presentation here will be more involved with the roots of the Dirac equation itself, rather than mathematical generality for its own sake.

In the first section of this study that follows this introduction, we shall discuss the way that the matrices in $M(2; \mathbb{C})$ can be used in order to represent non-relativistic and relativistic rotations (i.e., $SU(2)$ and $SL(2; \mathbb{C})$), as well as how that relates to Pauli spinors and Dirac bispinors. After that, we briefly summarize the Pauli extension of the Schrödinger equation for a charged particle in an external electromagnetic field to include a coupling of its magnetic moment to the magnetic field. We then review the relevant facts concerning Dirac's relativistic formulation of that picture, and pay particular attention to the role of duality in decomposing Dirac bispinors into pairs of Pauli spinors in orthogonal subspaces of $\mathbb{C}^4$. This then leads to the reformulation of the Dirac system of equations as a pair of first-order partial differential equations in the Pauli spinors, and by applying the "conjugate" of the minimally-coupled Dirac operator, one obtains a pair of second-order equations in two Pauli spinors, which can be combined into a single second-order equation for a complex 2×2 matrix-valued wave function. Along the way, we show how the basic transformations of Dirac bispinors, namely, Lorentz transformations, parity, time-reversal, and charge conjugation can be represented as transformations of matrices in $M(2; \mathbb{C})$. We conclude by presenting a Lagrangian for what we are calling the relativistic Pauli equation.

**2. The representation of spinors by matrices.** The complex vector space $M(2; \mathbb{C})$ of 2×2 complex matrices can be used to represent the vectors in $\mathbb{C}^4$, since they are isomorphic as complex vector spaces, but it also has an algebraic structure that is defined by matrix multiplication. In fact, when one gives $\mathbb{C}^4$ the algebra of complex quaternions, one finds that the two algebras are isomorphic. Hence, one could just as well deal with complex quaternions directly, but our reason for focusing on the matrix representation of that algebra is that it gives a convenient intermediate step between the traditional representations of spinors and the more abstract methods of quaternion algebra. Thus, we shall content ourselves to show how Pauli and Dirac spinors can both be represented by 2×2 complex matrices in a natural way, as well as the associated wave equations, and only refer the reader to a following article on the use of quaternions in the field equations of physics.

*a. The algebra $M(2; \mathbb{C})$.* A 2×2 complex matrix $[\psi]$ defines a pair of vectors in $\mathbb{C}^2$ or $\mathbb{C}^{2*}$ by way of its columns or rows, respectively. Thus, the first issue to address is whether they are linearly dependent or not. This is equivalent to the question of whether the determinant of $[\psi]$ vanishes or not, respectively.

If the determinant vanishes then the matrix can be put into the forms:

$$[\varphi \mid \lambda\varphi], \quad [\varphi \mid 0], \quad [0 \mid \chi], \quad \begin{bmatrix} \varphi \\ \lambda\varphi \end{bmatrix}, \quad \begin{bmatrix} \varphi \\ 0 \end{bmatrix}, \quad \begin{bmatrix} 0 \\ \chi \end{bmatrix},$$

which are essentially equivalent under a change of basis for $\mathbb{C}^2$ or its dual $\mathbb{C}^{2*}$. We shall then refer to such matrices as *null matrices*, and we note that since the determinant is a homogeneous quadratic function in the matrix elements, its vanishing defines a three-complex-dimensional quadric hypersurface in $M(2; \mathbb{C})$ that contains a infinitude of two-complex-dimensional linear subspaces of $M(2; \mathbb{C})$ whose elements take the forms that we just detailed. Note that since $\mathbb{C}$, unlike $\mathbb{R}$, does not split into two disconnected components when one removes 0, one cannot say that $\mathbb{C}^* = \mathbb{C} - 0$ defines two disjoint connected components in $M(2; \mathbb{C})$, as it does in $M(2; \mathbb{R})$.

In the general case, the two columns of $[\psi]$ are linearly independent, and therefore define a complex 2-frame in $\mathbb{C}^2$. One can then factor out the determinant from $[\psi]$ and express it in the form:

$$[\psi] = (\det [\psi])^{1/2} [\hat{\psi}],$$

in which $[\hat{\psi}]$ belongs to $SL(2; \mathbb{C})$. Thus, the columns of $[\hat{\psi}]$ span a complex parallelepiped that has unit volume, and we have also exhibited the group $GL(2; \mathbb{C})$ as the direct product group $\mathbb{C}^* \times SL(2; \mathbb{C})$, where $\mathbb{C}^*$ is the multiplicative group of non-zero complex numbers.

One can choose a volume element for $\mathbb{C}^2$ in the form of a non-zero bivector $\mathbf{a} \wedge \mathbf{b}$; thus, the vectors $\mathbf{a}$ and $\mathbf{b}$ must not be collinear, which is to say they are linearly independent. Since $\mathbb{C}^2$ is two-dimensional, the space of bivectors on it is one-dimensional, so any other non-zero bivector $\mathbf{v} \wedge \mathbf{w}$ must be a non-zero complex scalar multiple of $\mathbf{a} \wedge \mathbf{b}$. One finds that when one changes the basis $\{\mathbf{a}, \mathbf{b}\}$ by means of an invertible complex-linear transformation $L$, the volume element $\mathbf{a} \wedge \mathbf{b}$ goes to $(\det L) \, \mathbf{a} \wedge \mathbf{b}$. Hence, the elements of $SL(2; \mathbb{C})$ can be characterized as the matrices of invertible linear transformations of $\mathbb{C}^2$ that preserve the volume element.

When one gives $\mathbb{C}^2$ the Hermitian inner product:

$$(\mathbf{z}, \mathbf{w}) = \delta_{ij} z^i w^{j*},$$

in which the components of the vectors are expressed with respect to a unitary frame, one can also identify matrices whose columns define a unitary 2-frame in $\mathbb{C}^2$, which are then unitary matrices. They can be given the form:

$$\begin{bmatrix} \varphi^1 & -\varphi^{2*} \\ \varphi^2 & \varphi^{1*} \end{bmatrix} = [\varphi \mid \varphi^c],$$

in which $\varphi$ is a unit vector and, as we shall see, the column vector $\varphi^c = i\sigma^2 \varphi^*$ amounts to the complex conjugate of $\varphi$, rotated through $\pi/2$. One notes that a matrix of the present form also has a determinant of positive one, so it belongs to $SU(2)$. In order to produce a unitary matrix with a more general complex determinant (which must still lie on the unit circle), one can use $e^{i\alpha}\phi^c$, in place of $\phi^c$.

Despite its definition in terms of complex matrices, the group *SU*(2) is not a complex Lie group, but a real one; for one thing, the diagonal elements of any unitary matrix must be real. Indeed, *SU*(2) has a real dimension of three, which makes it impossible to give it a complex structure, to begin with. Although it is not clear from the definition that the manifold of *SU*(2) is a real three-dimensional sphere, this is easy to show when one represents the group by unit real quaternions. We shall not do this here, but simply refer to other sources for that demonstration (cf., [**13**], for one).

It is often useful to define a basis for the four-dimensional complex vector space *M*(2; $\mathbb{C}$). Although perhaps the simplest one is given by the four matrices with one element equal to unity and the other ones equal to zero, nevertheless, the one that is most convenient for the immediate purposes is the one that Pauli chose:

$$\sigma_0 = \begin{bmatrix} 1 & 0 \\ 0 & 1 \end{bmatrix}, \quad \sigma_1 = \begin{bmatrix} 0 & 1 \\ 1 & 0 \end{bmatrix}, \quad \sigma_2 = \begin{bmatrix} 0 & -i \\ i & 0 \end{bmatrix}, \quad \sigma_3 = \begin{bmatrix} 1 & 0 \\ 0 & -1 \end{bmatrix}. \tag{2.1}$$

Thus, all of the basis elements are Hermitian matrices, and the last three have trace zero. They obey the following multiplication rules:

$$\sigma_0 \sigma_i = \sigma_i \sigma_0 = \sigma_i, \qquad \sigma_i \sigma_j = \delta_{ij} + i\varepsilon_{ijk}\sigma_k . \tag{2.2}$$

This is almost isomorphic to the multiplication table of the complex quaternions, except that one must first replace the Pauli matrices $\sigma_i$ with the skew-Hermitian matrices:

$$\tau_i = -i\sigma_i, \tag{2.3}$$

in order to make the signs consistent. Clearly, the identity matrix $\sigma_0$ defines the unity of the algebra, which is associative, but not commutative.

From the multiplication rules, one can also conclude that:

$$\{\sigma_i, \sigma_j\} = 2\delta_{ij}, \qquad [\sigma_i, \sigma_j] = 2i\varepsilon_{ijk}\sigma_k . \tag{2.4}$$

Hence, the three $\sigma_i$ can represent the Lie algebra $\mathfrak{su}(2)$, which consists of all skew-Hermitian 2×2 complex matrices with trace zero, and represents the infinitesimal generators of one-parameter subgroups of special unitary transformations on $\mathbb{C}^2$. However, once again one must replace the $\sigma_i$ with the $\tau_i$ that were defined in (2.3). For that representation, the components $a^i$ of a matrix $a^i\tau_i$ in $\mathfrak{su}(2)$ with respect to this basis will be real.

The factor 2 that appears in (2.4) is significant, since it relates to the way that angular velocity gets represented in *SU*(2).

The three $\tau_i$ can also represent the Lie algebra $\mathfrak{sl}(2; \mathbb{C})$, which consists of all 2×2 complex matrices with trace zero, and represents the infinitesimal generators of one-parameter subgroups $SL(2; \mathbb{C})$. For that representation, one simply notes that any matrix $\Lambda \in \mathfrak{gl}(2; \mathbb{C})$ can be polarized into an anti-Hermitian matrix and a Hermitian one:

$$\Lambda = \omega + v, \qquad \omega = \tfrac{1}{2}(\Lambda - \Lambda^\dagger), \qquad v = \tfrac{1}{2}(\Lambda + \Lambda^\dagger), \qquad (2.5)$$

Thus, $\omega \in \mathfrak{u}(2)$ and $v \in \mathfrak{h}(2)$, which will denote the (real) vector space of Hermitian matrices in $M(2; \mathbb{C})$; by restriction to the matrices with trace zero, one has $\omega \in \mathfrak{su}(2)$ and $v \in \mathfrak{sh}(2)$. Since the matrices $\tau_i$ define a real basis for $\mathfrak{su}(2)$ and the matrices $i\tau_i$ define an imaginary basis for $\mathfrak{sh}(2)$, the $\tau_i$ define a complex basis for $\mathfrak{sl}(2; \mathbb{C})$ that makes the general matrix $\Lambda \in \mathfrak{sl}(2; \mathbb{C})$ take the form:

$$\Lambda = (\omega^i + iv^i)\tau_i. \qquad (2.6)$$

In order to go from the representation of infinitesimal rotations or Lorentz transformations by matrices in $M(2; \mathbb{C})$ to the representation of finite ones, we shall simply use the exponential map, as it is defined for square matrices:

$$\exp[M] = \sum_{n=0}^{\infty} \frac{1}{n!}[M]^n. \qquad (2.7)$$

Every element $a$ of $\mathfrak{su}(2)$ or $\mathfrak{sl}(2; \mathbb{C})$ then defines a one-parameter subgroup of $SU(2)$ or $SL(2; \mathbb{C})$, resp., by way of $\exp(sa)$. As long as one only considers the identity components of a matrix Lie group, all of its elements can be represented this way.

Although the anti-commutation rules in (2.4) look identical to those of the Clifford algebra over real three-dimensional Euclidian space, this is misleading, since that Clifford algebra has a real dimension of eight and if $\{\mathbf{e}_1, \mathbf{e}_2, \mathbf{e}_3\}$ is an orthonormal basis for it then the Clifford algebra has a basis defined by all of the independent products of these basis elements when one imposes the constraint that $\mathbf{e}_i \mathbf{e}_j + \mathbf{e}_j \mathbf{e}_i = 2\delta_{ij}$. However, this means that $\mathbf{e}_i \mathbf{e}_j$ is linearly independent of all three of the $\mathbf{e}_i$, so it one associates the basis elements with the Pauli matrices then the products become inconsistent; i.e., the association of basis elements does not generate an algebra homomorphism.

*b. Pauli spinors.* The Pauli spinors are wave functions $\varphi(t, x^i)$ on spacetime that take their values in $\mathbb{C}^2$ or $\mathbb{C}^{2*}$ and describe the non-relativistic motion of spinning particles. Thus, the group that pertains to the rotational motion is $SU(2)$, which is the double covering group for $SO(3)$.

If one thinks of the elements of $\mathbb{C}^2$ as column vectors then $SU(2)$ acts on $\mathbb{C}^2$ by matrix multiplication on the left. If $\mathbb{C}^{2*}$ is represented by row vectors then it also acts on the dual space $\mathbb{C}^{2*}$ by matrix multiplication on the right. However, in order to be consistent with the inner product, one might also make it act on the right by Hermitian adjoint. That is, if $\alpha \in \mathbb{C}^{2*}$ is a row vector and $U \in SU(2)$ is a special unitary matrix then one can define:

$$(\alpha, U) = \alpha U^\dagger.$$

This would make the bilinear pairing:

$$<\alpha, \mathbf{v}> = \alpha(\mathbf{v}) = \boldsymbol{\alpha}^\dagger \mathbf{v}$$

invariant under the left action of $SU(2)$ on $\mathbb{C}^2$ and its right action by Hermitian adjoint on $\mathbb{C}^{2*}$. Here, we are representing the row vector $\alpha \in \mathbb{C}^{2*}$ by means of the Hermitian conjugate of a column vector $\boldsymbol{\alpha} \in \mathbb{C}^2$.

$SU(2)$ also acts on $M(2; \mathbb{C})$ in various ways. Of particular interest here is the action:

$$SU(2) \times M(2; \mathbb{C}) \to M(2; \mathbb{C}), \ (U, M) \mapsto UMU^\dagger. \tag{2.8}$$

This action is the one that allows us to relate $SU(2)$ to real proper Euclidian rotations in three dimensions. This is because the association of an orthonormal basis $\{\mathbf{e}_1, \mathbf{e}_2, \mathbf{e}_3\}$ for $\mathbb{R}^3$ with the basis $\{\sigma_1, \sigma_2, \sigma_3\}$ defines an $\mathbb{R}$-linear isomorphism of $\mathbb{R}^3$ with a three-dimensional real subspace of $\mathbb{C}^2$, namely, the one that is occupied by the Hermitian matrices, which we will denote by $H(2)$. Note that $H(2)$ is not a subalgebra of $M(2; \mathbb{C})$, since the product of two Hermitian matrices does not have to be Hermitian.

If a vector $\mathbf{v} \in \mathbb{R}^3$ is represented by $v^i \mathbf{e}_i$ then its isomorphic image in $H(2)$ is the Hermitian matrix:

$$v^i \sigma_i = \begin{bmatrix} v^3 & v^1 - iv^2 \\ v^1 + iv^2 & -v^3 \end{bmatrix}. \tag{2.9}$$

Note that the determinant of the matrix $v^i \sigma_i$ takes a particularly, useful form:

$$\det(v^i \sigma_i) = - \| \mathbf{v} \|^2. \tag{2.10}$$

(The minus sign is due to the choice of Hermitian matrices over skew-Hermitian ones.)

One can give $H(2)$ a Euclidian scalar product by way of the Cartan-Killing form:

$$<H_1, H_2> = \text{Tr}(H_1 H_2). \tag{2.11}$$

One then sees that the Pauli basis is orthonormal for this scalar product:

$$<\sigma_i, \sigma_j> = \delta_{ij}. \tag{2.12}$$

This also means that the vector space of Hermitian matrices is isometric to three-dimensional real Euclidian space under the present association of basis elements. Hence, it can essentially represent the geometry of that space, and, in particular, its rotations.

If we go back to the action (2.8) then we see that it takes Hermitian matrices to other Hermitian matrices, since if $H$ is Hermitian and $U$ is unitary then:

$$(UHU^\dagger)^\dagger = UH^\dagger U^\dagger = UHU^\dagger, \tag{2.13}$$

in which we have implicitly used the properties of the Hermitian dagger that amount to saying that is it an anti-involution of the algebra $M(2; \mathbb{C})$, namely:

$$(M^\dagger)^\dagger = M, \qquad (AB)^\dagger = B^\dagger A^\dagger. \tag{2.14}$$

Since this action is linear, it can also be represented by a linear map $L(U): H(2) \to H(2)$, $M \mapsto UMU^\dagger$, so relative to the basis of $\sigma_i$ it can be described by a 3×3 matrix $[L(U)]_j^i$, namely:

$$U\sigma_i U^\dagger = \sigma_j \, [L(U)]_i^j. \tag{2.15}$$

Note that the matrices $\sigma_j$ and $[L(U)]_i^j$ cannot possibly multiply in this expression, so one must regard the matrices $\sigma_j$ as simply abstract vectors and the elements of the matrix $[L(U)]_i^j$ are scalar coefficients that multiply them.

One sees that both $U$ and $-U$ produce the same $L(U)$, due to the quadratic nature of the action. Furthermore, the action of $SU(2)$ on $H(2)$ preserves the Euclidian scalar product that is defined by the Cartan Killing form since:

$$<L(U)H_1, L(U)H_2> = \text{Tr}(UH_1 U^\dagger UH_2 U^\dagger) = \text{Tr}(UH_1 H_2 U^\dagger) = \text{Tr}(H_1 H_2) = <H_1, H_2>.$$

The matrix $[L(U)]_i^j$ is also real, since it must take Hermitian matrices to Hermitian matrices. Thus, when the components of $[L(U)]_i^j$ as scalars that multiply the vectors $\sigma_j$, we get:

$$(\sigma_j \, [L(U)]_i^j)^\dagger = \sigma_j^\dagger [L(U)]_i^{j*} = \sigma_i \, [L(U)]_i^{j*} = \sigma_j \, [L(U)]_i^j,$$

which then implies that $[L(U)]_i^j = [L(U)]_i^{j*}$.

So far, we have shown that $[L(U)]_i^j$ is in $O(3; \mathbb{R})$. One can show that it also has unity determinant, although if one knows that the manifold that $U$ comes from is connected then its image under the continuous map $L: SU(2) \to GL(2; \mathbb{C})$, $U \mapsto L(U)$ must also be connected, and the connected component of $O(3; \mathbb{R})$ is the one that contains the identity matrix, namely, $SO(3; \mathbb{R})$. We have thus exhibited the two-to-one covering homomorphism $SU(2) \to SO(3; \mathbb{R})$ within the context of the Euclidian space of Hermitian matrices.

A Pauli spinor is a wave function $\varphi(t, x^i) = [\varphi^1, \varphi^2]^T$ that takes its values in $\mathbb{C}^2$. There are two distinct way in which one can interpret the vector $\varphi(t, x^i)$ as a matrix in $M(2; \mathbb{C})$.

As we have seen, the null matrices in $M(2; \mathbb{C})$ include many subspaces that can represent $\mathbb{C}^2$ or its dual $\mathbb{C}^{2*}$. The ones that we shall use are:

$$[\varphi] = \begin{bmatrix} \varphi^1 & 0 \\ \varphi^2 & 0 \end{bmatrix}, \qquad [\varphi]^\dagger = \begin{bmatrix} \varphi^{1*} & \varphi^{2*} \\ 0 & 0 \end{bmatrix}. \tag{2.16}$$

The left or right action of $SU(2)$ on these matrices then duplicates its left or right action on the elements of $\mathbb{C}^2$ or $\mathbb{C}^{2*}$, respectively, if one ignores the trivial action on the null column or row, resp. Although this sounds gratuitous at the moment, we shall see that the transition from Pauli spinors to Dirac spinors becomes straightforward when one thinks of both of them as matrices.

One also finds that $SU(2)$ – and therefore the angular orientation of a wave – can be represented by points of $\mathbb{C}^2$, as well as by 2×2 special unitary matrices. Namely, if one has a unit vector $\varphi$ in $\mathbb{C}^2$, relative to the Hermitian inner product, then one can define a matrix in $SU(2)$ by way of:

$$\begin{bmatrix} \varphi^1 & -\varphi^{2*} \\ \varphi^2 & \varphi^{1*} \end{bmatrix}.$$

Clearly, it has unity determinant, so it is invertible, and its inverse is also seen to be its Hermitian conjugate; thus, every unit vector in $\mathbb{C}^2$ defines an element of $SU(2)$. One should note that the fact that any vector in $\mathbb{C}^2$ defines a rotation in $SU(2)$, as well as an oriented unitary frame, which we shall refer to as a *Pauli spin frame*, is closely analogous to the way that any vector in the real plane defines a real rotation and an oriented orthonormal frame. Thus, in a sense, a Pauli spinor is an "abbreviation" for a Pauli spin frame.

One should note, for later reference, that the second column takes the form:

$$\begin{bmatrix} -\varphi^{2*} \\ \varphi^{1*} \end{bmatrix} = \begin{bmatrix} 0 & -1 \\ 1 & 0 \end{bmatrix} \begin{bmatrix} \varphi^{1*} \\ \varphi^{2*} \end{bmatrix} = -i\sigma^2 \varphi^*,$$

which will prove relevant to the context of charge conjugation..

*c. Dirac bispinors.* A Dirac bispinor is a wave function $\psi(t, x^i)$ on spacetime that takes its values in $\mathbb{C}^4$, although generally the values can also be treated as order pairs of Pauli spinors:

$$\psi = \begin{bmatrix} \psi^1 \\ \psi^2 \\ \psi^3 \\ \psi^4 \end{bmatrix} = \begin{bmatrix} \psi_u \\ \psi_d \end{bmatrix}.$$

The linear transformations of the field space can then be represented by matrices in $M(4; \mathbb{C})$. As a 16-complex-dimensional matrix algebra, it is isomorphic to the complex Clifford algebra $\mathcal{C}_\mathbb{C}(4; \eta_{\mu\nu})$ that is generated by complex Minkowski space $\mathfrak{M}_\mathbb{C}^4$. Thus, if $\{\mathbf{e}_\mu, \mu = 0, \ldots, 3\}$ is an orthonormal basis for $\mathfrak{M}_\mathbb{C}^4$ then the products of all the basis elements must satisfy:

$$\mathbf{e}_\mu \mathbf{e}_\nu + \mathbf{e}_\nu \mathbf{e}_\mu = 2\eta_{\mu\nu}. \qquad (2.17)$$

The remaining twelve basis elements of $\mathcal{C}_\mathbb{C}(4; \eta_{\mu\nu})$ are then defined by 1 and the linearly independent products of the form $\mathbf{e}_\mu \mathbf{e}_\nu$, $\mathbf{e}_\lambda \mathbf{e}_\mu \mathbf{e}_\nu$, and $\mathbf{e}_0 \mathbf{e}_1 \mathbf{e}_2 \mathbf{e}_3$.

Of course, it is actually *real* Minkowski space that is most relevant to relativistic motion, and one represents its Clifford algebra by means of the real sub-algebra of $\mathcal{C}_\mathbb{C}(4; \eta_{\mu\nu})$ that is spanned by the sixteen basis elements when they are given real components. Thus, in a sense, the choice of $\mathcal{C}_\mathbb{C}(4; \eta_{\mu\nu})$ as a representation space is excessive by a factor of two.

The basis elements for $M(4; \mathbb{C})$ that one uses for the Dirac equation are generated by four matrices $\gamma_\mu$, $\mu = 0, \ldots, 3$ that one associates with the orthonormal basis defined by the $\mathbf{e}_\mu$. Thus, they must satisfy the basic relation:

$$\gamma_\mu \gamma_\nu + \gamma_\nu \gamma_\mu = 2\eta_{\mu\nu} I, \qquad (2.18)$$

in which $I$ is the 4×4 identity matrix.

A particular fundamental matrix that one forms from the gamma matrices is:

$$\gamma^5 = \gamma^0 \gamma^1 \gamma^2 \gamma^3,$$

which has much the same effect on products of the gamma matrices as the Hodge duality operator does on exterior $k$-forms on $\mathbb{R}^4$.

One realizes that just as any orthonormal basis for $\mathcal{C}(4; \eta^{\mu\nu})$ will satisfy the relations (4.2), and will thus serve as a set of generators, similarly there will be an infinitude of representations for the gamma matrices. Three of the most popular ones are the Dirac, Weyl, and Majorana representations, all of which have in common that they are based in the Pauli matrices as blocks in the matrices and act upon bispinors.

The Dirac representation, which is also used by Bjørken and Drell [**14**], defines:

$$\gamma^0 = \begin{bmatrix} I & 0 \\ 0 & -I \end{bmatrix}, \qquad \gamma^i = \begin{bmatrix} 0 & \sigma^i \\ -\sigma^i & 0 \end{bmatrix}, \qquad (2.19)$$

which makes:

$$i\gamma^5 = -\begin{bmatrix} 0 & I \\ I & 0 \end{bmatrix}. \qquad (2.20)$$

Thus, the operator $i\gamma^5$ has the useful property that it simply permutes $\psi_u$ and $\psi_d$, while changing their signs.

The Weyl representation defines:

$$\gamma^0 = \begin{bmatrix} 0 & I \\ I & 0 \end{bmatrix}, \qquad \gamma^i = \begin{bmatrix} 0 & -\sigma^i \\ \sigma^i & 0 \end{bmatrix}, \qquad (2.21)$$

which makes:

$$\gamma^5 = \begin{bmatrix} I & 0 \\ 0 & -I \end{bmatrix}. \qquad (2.22)$$

Thus, the roles of $\gamma^0$ and $\gamma^5$ seem to have been switched, while the only change to the spatial matrices is their sign.

One sees that the Weyl representation has diagonalized the matrix $\gamma^5$, whose role we will discuss in more detail in the next subsection. One finds that the unitary matrix $U$ that converts the Dirac matrices to Weyl matrices by way of $U\gamma U^\dagger$ is:

$$U = \frac{1}{\sqrt{2}} \begin{bmatrix} I & I \\ I & -I \end{bmatrix}. \qquad (2.23)$$

What Majorana [**11**] was looking for was a way to represent neutrino wave functions, and since neutrinos are uncharged, under the charge conjugation operator they should be self-conjugate. Thus, he thought it would be particularly useful if the charge conjugation operator took the form of complex conjugation of the wave function, so the uncharged particles would be represented by real wave functions. The representation that he chose was then:

$$\gamma^0 = \begin{bmatrix} 0 & \sigma^2 \\ \sigma^2 & 0 \end{bmatrix}, \quad \gamma^1 = \begin{bmatrix} i\sigma^3 & 0 \\ 0 & i\sigma^3 \end{bmatrix}, \quad \gamma^2 = \begin{bmatrix} 0 & -\sigma^2 \\ \sigma^2 & 0 \end{bmatrix}, \quad \gamma^3 = \begin{bmatrix} -i\sigma^1 & 0 \\ 0 & -i\sigma^1 \end{bmatrix}, \quad (2.24)$$

which makes:

$$\gamma^5 = \begin{bmatrix} \sigma^2 & 0 \\ 0 & -\sigma^2 \end{bmatrix}. \quad (2.25)$$

One finds that all of the matrices above are pure imaginary, so one always has:

$$\gamma^{\mu *} = -\gamma^{\mu}. \quad (2.26)$$

When we get to the coupling of the Dirac wave function to an external electromagnetic field, we shall see how this simplifies the charge conjugation operator.

*d. Duality for Dirac bispinors.* One can use the $i\gamma^5$ matrix, in the case of the Dirac representation (or $\gamma^5$, in the Weyl representation) to define projection operators $P_\pm$ that split $\mathbb{C}^4$ into a direct sum $\Sigma_+ \oplus \Sigma_-$ of two-dimensional subspaces that amount to the eigenspaces of the matrix $i\gamma^5$ ($\gamma^5$, resp.). In either case, the eigenvalues are $\pm 1$, so we shall refer to the elements of $\Sigma_+$ ($\Sigma_-$, resp.) as *self-dual* (*anti-self-dual,* resp.) bispinors.

This direct sum decomposition also entails a decomposition of the identity operator into a sum of complementary projection operators:

$$I = P_+ + P_- . \quad (2.27)$$

For the Dirac case, the eigenvalue equation for the duality operator is:

$$i\gamma^5 \psi = \begin{bmatrix} -\psi_d \\ -\psi_u \end{bmatrix} = \begin{bmatrix} \pm \psi_u \\ \pm \psi_d \end{bmatrix},$$

so the eigenvectors take the form:

$$(+) = \begin{bmatrix} \varphi \\ -\varphi \end{bmatrix}, \qquad (-) = \begin{bmatrix} \chi \\ \chi \end{bmatrix}.$$

and the projection operators take the form:

$$P_\pm = \tfrac{1}{2}(I \pm i\gamma^5), \qquad P_+ = \tfrac{1}{2}\begin{bmatrix} 1 & -1 \\ -1 & 1 \end{bmatrix}, \qquad P_- = \tfrac{1}{2}\begin{bmatrix} 1 & 1 \\ 1 & 1 \end{bmatrix}, \quad (2.28)$$

These eigenvectors are, moreover, orthogonal for the Hermitian inner product:

$$(+)^\dagger(-) = (-)^\dagger(+) = 0.$$

Hence, the subspaces $\Sigma_+$ and $\Sigma_-$ are orthogonal subspaces of $\mathbb{C}^4$.

For the Weyl case, the eigenvalue equation takes the form:

$$\gamma^5 \psi = \begin{bmatrix} \psi_u \\ -\psi_d \end{bmatrix} = \begin{bmatrix} \pm \psi_u \\ \pm \psi_d \end{bmatrix},$$

so the eigenvectors take the form:

$$(+) = \begin{bmatrix} \varphi \\ 0 \end{bmatrix}, \qquad (-) = \begin{bmatrix} 0 \\ \chi \end{bmatrix}.$$

and the projections operators become:

$$P_\pm = \tfrac{1}{2}(I \pm \gamma^5), \qquad P_+ = \begin{bmatrix} 1 & 0 \\ 0 & 0 \end{bmatrix}, \qquad P_- = \begin{bmatrix} 0 & 0 \\ 0 & 1 \end{bmatrix}. \qquad (2.29)$$

Thus, in the Weyl case, the direct sum represents simply the sum of the up and down spaces. In fact, the unitary matrix $U$ that takes the Dirac gamma matrices to the Weyl gamma matrices is effectively the matrix that will diagonalize the Dirac matrix $i\gamma^5$ into the Weyl matrix $-\gamma^5$.

From now on, we shall always denote the "up" and "down" spinors of a bispinor in Weyl form by $\varphi$ and $\chi$, respectively.

In either the Dirac or Weyl representation, the operator $\gamma^0$ then defines a $\mathbb{C}$-linear unitary isomorphism $\gamma^0 : \Sigma_+ \to \Sigma_-$, and its inverse is also $\gamma^0$. As it happens, this transformation also relates to a basic physical symmetry of quantum wave functions, namely, parity. One finds (cf., Bjørken and Drell [14]) that the spatial inversion operator, which takes $(t, x^i)$ to $(t, -x^i)$, gets represented in the field space of Dirac wave functions by the operator $P$ that makes:

$$\psi^P(t, -x^i) = P\psi(t, x^i),$$

namely:

$$P = e^{i\phi} \gamma^0, \qquad (2.30)$$

in which the phase $\phi$ can be chosen to suit one's purposes. Thus, the parity opposite of a self-dual wave function is an anti-self-dual one, and vice versa.

Since the even and odd wave functions will be the ones for which $\psi^P(t, -x^i) = \pm \psi(t, x^i)$, resp., they will be the ones for which $\psi$ take the forms:

$$\psi = \begin{bmatrix} \psi_u \\ 0 \end{bmatrix} \quad \text{or} \quad \begin{bmatrix} 0 \\ \psi_d \end{bmatrix}, \text{ resp.},$$

in which $\psi_u$ will be an even function and $\psi_d$ will be an odd function, in the usual sense. Thus, the only self-dual or anti-self-dual even or odd wave function is zero, and all other even or odd wave functions must be expressed as linear combinations of self-dual and anti-self-dual wave functions.

Note that in the Weyl representation, the roles of parity and duality seem to get reversed.

*e. Representation of the proper, orthochronous Lorentz group in $\mathbb{C}^2$.* If $(v^0, v^1, v^2, v^3)$ is a vector in $\mathbb{R}^4$ then one can represent it in $M(2; \mathbb{C})$ by the 2×2 complex Hermitian matrix:

$$[\mathbf{v}] = v^\mu \sigma_\mu = \begin{bmatrix} v^0 + v^3 & v^1 - iv^2 \\ v^1 + iv^2 & v^0 - v^3 \end{bmatrix}. \tag{2.31}$$

Thus, the four-real-dimensional subspace of $M(2; \mathbb{C})$ that represents $\mathbb{R}^4$ is $H(2)$.

If one takes the determinant of the matrix $[\mathbf{v}]$ then one gets:

$$\det[\mathbf{v}] = \eta_{\mu\nu} v^\mu v^\nu, \tag{2.32}$$

which says that the determinant plays the same role in $M(2; \mathbb{C})$ that the Minkowski scalar product does in Minkowski space. Thus, linear transformations of $M(2; \mathbb{C})$ that preserve the determinant – i.e., elements of $SL(2; \mathbb{C})$ – effectively preserve the Minkowski scalar product, which is then defined by:

$$\langle [\mathbf{v}], [\mathbf{w}] \rangle = \det([\mathbf{v}] [\mathbf{w}]^\dagger).$$

However, the action of $SL(2; \mathbb{C})$ on $M(2; \mathbb{C})$ that corresponds to the action of the proper, orthochronous Lorentz group $SO_+(3; 1)$ on Minkowski space is that of conjugation; namely, if $T \in SL(2; \mathbb{C})$ then the matrix $[\mathbf{v}]$ goes to $T[\mathbf{v}]T^\dagger$, which is easily seen to be Hermitian, as well.

Since this action is linear on $M(2; \mathbb{C})$, it can also be represented by a 4×4 matrix $L_\mu^\nu$:

$$T \sigma_\mu T^\dagger = \sigma_\mu L_\mu^\nu. \tag{2.33}$$

Since the $\sigma_\mu$ are Hermitian, the matrix $L_\mu^\nu$ is seen to be real by taking the Hermitian conjugate of both sides of (2.33), and keeping in mind that the matrix $L_\mu^\nu$ is treated as a set of scalar multipliers under this operation:

$$(T \sigma_\mu T^\dagger)^\dagger = \sigma_\mu L_\mu^{\nu *} = T \sigma_\mu T^\dagger = \sigma_\mu L_\mu^\nu,$$

which then implies that:

$$L_\mu^{\nu *} = L_\mu^\nu.$$

We can also define the action of $L_\mu^\nu$ on $H(2)$ by its action on $\mathbb{R}^4$:

$$L[\mathbf{v}] = [L\mathbf{v}] = (L_\nu^\mu v^\nu)\sigma_\mu,$$

and one finds that

$$\langle[L\mathbf{v}], [L\mathbf{w}]\rangle = \langle L\mathbf{v}, L\mathbf{w}\rangle.$$

Furthermore, one finds that $L_\mu^\nu$ must preserve the Minkowski scalar product on $\mathbb{R}^4$, since:

$$\langle T[\mathbf{v}]T^\dagger, T[\mathbf{w}]T^\dagger\rangle = \det(T[\mathbf{v}]T^\dagger T[\mathbf{w}]^\dagger T^\dagger) = \det([\mathbf{v}][\mathbf{w}]^\dagger) = \langle[\mathbf{v}], [\mathbf{w}]\rangle$$

implies that

$$\langle[L\mathbf{v}], [L\mathbf{w}]\rangle = \langle[\mathbf{v}], [\mathbf{w}]\rangle.$$

Thus, we have a map $SL(2; \mathbb{C}) \to SO_+(3; 1)$, $T \mapsto L$, which is seen to be two-to-one, since $T$ and $-T$ both give the same transformation of $H(2)$ under conjugation. In fact, this map is a group homomorphism that represents the covering map of the identity component of $SO_+(3; 1)$.

*f. Dirac bispinors as matrices.* We shall find that the Weyl representation of the gamma matrices is most convenient for most of our purposes. In particular, when one associates the Dirac bispinor in Weyl form $\psi = [\varphi, \chi]^T$ with the 2×2 complex matrix:

$$[\psi] = [\varphi \,|\, \chi] \qquad (2.34)$$

the Dirac equation will be converted into a form in which the introduction of the gamma matrices and a choice of representation for them will no longer be necessary. Furthermore, the somewhat artificial representation of the Lorentz group by 4×4 complex matrices that act on Dirac bispinors gets replaced with the more elementary representation in $SL(2; \mathbb{C})$, which acts on the matrices $[\psi]$ either on the left or right, depending upon how the matrix is interpreted.

Thus, a Dirac bispinor is still a pair of Pauli spinors, but the second one should be interpreted as a relativistic correction to the non-relativistic spinor that takes into account the boost transformations, much as one speaks of the "large" and "small" components of a Dirac bispinor. The null case of a matrix with zero determinant then represents a Pauli spinor again, so in the non-null case the columns of $[\psi]$ define a *Dirac spin frame* in $\mathbb{C}^2$, while the rows define a Dirac spin frame in $\mathbb{C}^{2*}$. In the event that the spin frame is unitary, one comes back to Pauli spinors, as well as when on row or column vanishes.

**3. The Pauli electron.** In order to model the spin of the non-relativistic electron, Pauli chose the space $\mathbb{C}^2$ for the field space of its wave function. Having discussed the representation of rotations by way of the action of 2×2 special unitary matrices on 2×2 Hermitian matrices and their representation as Pauli spinors, we now simply present the equation that Pauli proposed as the extension of the Schrödinger equation to spinning electrons in an external electromagnetic field.

Let $\varphi$ represent the Pauli spinor wave function of the electron, let $e$ be its electric charge, and let $m$ be its mass. Then, let $\phi$ be the electrostatic potential of the external field and let **A** be the vector potential of the external magnetic field $\mathbf{B} = \nabla \times \mathbf{A}$. One first minimally couples the vector potential to the momentum by the usual prescription $\mathbf{p} \to \mathbf{p} - (e/c)\mathbf{A}$ and adds the potential energy $e\phi$ of the charge in the external electrostatic field as a potential term in the Schrödinger equation, and all that is missing is the term that couples the magnetic moment of the electron to the external magnetic field, which takes the form of the potential energy due to the torque produced:

$$\boldsymbol{\mu} \cdot \mathbf{B} = -\frac{e\hbar}{2mc} B^i \sigma_i. \tag{3.1}$$

The ultimate equation is:

$$i\hbar \frac{\partial \varphi}{\partial t} = \left[ \frac{1}{2m}\left( i\hbar \frac{\partial}{\partial x^i} - \frac{e}{c} A_i \right)^2 + e\phi - \frac{e\hbar}{2mc} B^i \sigma_i \right] \varphi. \tag{3.2}$$

One then sees that if it were not for the fact that the matrix $B^i \sigma_i$ is not generally diagonal there would be no coupling of the two components of $\varphi$, and the system would degenerate to an independent pair of equations for the individual components. However, if **B** points in the $z$-direction then the matrix $B^i \sigma_i = B^3 \sigma_3$ *will* be diagonal for this choice of basis on $H(2)$.

**4. The Dirac electron.** Dirac's way of finding a square root to the Klein-Gordon operator was to formally start with an operator of the form:

$$i\hbar(\partial_0 + \alpha^i \partial_i) - mc\beta \qquad (x^0 = ct)$$

whose coefficients $\alpha^i, \beta$ were not necessarily scalars, but would still commute with the partial derivative operators, and multiply it by its "conjugate":

$$[i\hbar(\partial_0 + \alpha^i \partial_i) - mc\beta][i\hbar(-\partial_0 + \alpha^i \partial_i) - mc\beta]$$
$$= \hbar^2 (\partial_0^2 - \alpha^i \alpha^j \partial_i \partial_j) - i\hbar mc(\alpha^i \beta + \beta \alpha^i)\partial_i + m^2 c^2 \beta^2.$$

In order for this to equal to the Klein-Gordon operator:

$$\hbar^2 \eta^{\mu\nu} \partial_\mu \partial_\nu + m^2 c^2 \qquad (\eta^{\mu\nu} = \text{diag}[1, -1, -1, -1])$$

one would have to have:

$$\alpha^i \alpha^j + \alpha^j \alpha^i = 2\delta^{ij}, \qquad \alpha^i \beta + \beta \alpha^i = 0, \qquad \beta^2 = 1.$$

The first set of these relations suggest that the $\alpha^i$ belong to the Clifford algebra of three-dimensional real Euclidian space, which still sounds non-relativistic. The last relation says that $\beta$ is invertible, and in fact, equal to its own inverse. Hence, one can left-multiply the original first order operator by that matrix to put it into the form:

$$i\hbar \gamma^\mu \partial_\mu - mc, \qquad (\gamma^0 = \beta,\ \gamma^i = \beta \alpha^i),$$

or, if we divide through by $i\hbar$ then we get the Dirac operator in the form:

$$\slashed{\partial} + i\kappa = \gamma^\mu \partial_\mu + i\kappa,$$

into which we have now introduced the Compton wave number for the particle $\kappa = mc/\hbar$.

*a. The Dirac equation.* This gives us the Dirac equation, which we express in the form:

$$\slashed{\partial}\psi = -i\kappa\psi. \tag{4.1}$$

If we define the conjugate to the operator $\slashed{\partial} + i\kappa$ to be $\slashed{\partial} - i\kappa$ then when we multiply the two (in either order) we get:

$$(\gamma^\mu \partial_\mu - i\kappa)(\gamma^\mu \partial_\mu + i\kappa) = \gamma^\mu \gamma^\nu \partial_{\mu\nu} + \kappa^2 = \tfrac{1}{2}(\gamma^\mu \gamma^\nu + \gamma^\nu \gamma^\mu)\partial_{\mu\nu} + \kappa^2.$$

In order to give the Klein-Gordon operator, the coefficients $\gamma^\mu$ must now satisfy:

$$\gamma^\mu \gamma^\nu + \gamma^\nu \gamma^\mu = 2\eta^{\mu\nu} \qquad (\eta^{\mu\nu} = \text{diag}[+1, -1, -1, -1]), \tag{4.2}$$

which defines the Clifford algebra $\mathcal{C}(4;\eta^{\mu\nu})$ of real Minkowski space if one assumes that the coefficients $\gamma^\mu$, $\mu = 0, \ldots, 3$ represent an orthonormal basis for it.

When one uses the Dirac or Weyl representations of the gamma matrices, since both representations take the form of block matrices that involve the Pauli matrices (including the identity matrix), one can also present the Dirac equation as a pair of equations for the Pauli spinors $\psi_u$ and $\psi_d$.

With the Dirac representation, the operator $\slashed{\partial}$ takes the form:

$$\slashed{\partial} = \begin{bmatrix} \partial_0 \psi_u + \sigma^i \partial_i \psi_d \\ -\partial_0 \psi_d - \sigma^i \partial_i \psi_u \end{bmatrix}. \tag{4.3}$$

so the Dirac equation takes the form of the following system:

$$\begin{cases} \dfrac{\partial \psi_u}{\partial x^0} + \sigma^i \dfrac{\partial \psi_d}{\partial x^i} = -i\kappa \psi_u, \\ \dfrac{\partial \psi_d}{\partial x^0} + \sigma^i \dfrac{\partial \psi_u}{\partial x^i} = +i\kappa \psi_d. \end{cases} \quad (4.4)$$

In this form, one sees that the up and down Pauli spinors are coupled by way of their spatial derivatives.

With the Weyl representation, the operator $\partial\!\!\!/$ takes the form:

$$\partial\!\!\!/ = \begin{bmatrix} \partial_0 \psi_d - \sigma^i \partial_i \psi_d \\ \partial_0 \psi_u + \sigma^i \partial_i \psi_u \end{bmatrix}, \quad (4.5)$$

and the Dirac equation decomposes into:

$$\begin{cases} \dfrac{\partial \psi_u}{\partial x^0} + \sigma^i \dfrac{\partial \psi_u}{\partial x^i} = -i\kappa \psi_d, \\ \dfrac{\partial \psi_d}{\partial x^0} - \sigma^i \dfrac{\partial \psi_d}{\partial x^i} = -i\kappa \psi_u. \end{cases} \quad (4.6)$$

One still has a coupling of the up and down spinors, although not by way of the differential operators.

One sees that if we define the differential operators:

$$\nabla = \dfrac{\partial}{\partial x^0} + \sigma^i \dfrac{\partial}{\partial x^i}, \qquad \overline{\nabla} = \dfrac{\partial}{\partial x^0} - \sigma^i \dfrac{\partial}{\partial x^i} \quad (4.7)$$

then the Weyl form (4.6) of the Dirac equation can be put into the concise form:

$$\nabla \psi_u = -i\kappa \psi_d, \qquad \overline{\nabla} \psi_d = -i\kappa \psi_u. \quad (4.8)$$

*b. The Dirac equation for self-dual and anti-self-dual bispinors.* Although the Dirac projection operators appear to be less geometrically intuitive, nevertheless, they prove to be more useful in decoupling the Dirac equation into a pair of first-order wave equations.

We see that the effect of the operator $\partial\!\!\!/$ on self-dual wave functions of the form $[\varphi, -\varphi]^T$ is:

$$\partial\!\!\!/ \psi_+ = \begin{bmatrix} \overline{\nabla} \varphi \\ \overline{\nabla} \varphi \end{bmatrix}, \quad (4.9)$$

so the result is an anti-self-dual wave function; that is, $\partial\!\!\!/ : \Sigma_+ \to \Sigma_-$.

Since the wave function:

$$-i\kappa \psi_+ = \begin{bmatrix} -i\kappa \varphi \\ i\kappa \varphi \end{bmatrix}$$

still takes its values in $\Sigma_+$, we see that the Dirac equation can admit such solutions in a non-trivial way only if $\kappa = 0$; i.e., they must be massless waves. For such waves, the Dirac equation reduces to the following equation for the Pauli spinor $\varphi$:

$$\bar{\nabla}\varphi = 0. \tag{4.10}$$

One finds that an analogous situation prevails for the anti-self-dual wave functions. Since they have the form $\psi_- = [\chi, \chi]^T$, we see that the effect of the operator $\partial\!\!\!/$ on such a wave function is to produce a wave function of the form:

$$\partial\!\!\!/\,\psi_- = \begin{bmatrix} \nabla\chi \\ -\nabla\chi \end{bmatrix}, \tag{4.11}$$

which is then self-dual; thus, $\partial\!\!\!/ : \Sigma_- \to \Sigma_+$. Therefore, since $-i\kappa\psi_-$ must still belong to $\Sigma_-$, the only way that the Dirac equation can admit non-trivial solutions for $\psi_-$ is, once more, if $\kappa = 0$; i.e., they must be massless waves, for which the Dirac equation becomes:

$$\nabla\chi = 0. \tag{4.12}$$

In summary, the only self-dual or anti-self-dual solutions to the Dirac equations are massless waves. However, since the Dirac bispinor is the sum of a self-dual and an anti-self-dual spinor, the transforms of the individual Pauli spinors can still couple to the opposite spinors. That is, if one expresses the Dirac equation in the form:

$$\left.\begin{aligned}\frac{\partial\varphi}{\partial x^0}+\frac{\partial\chi}{\partial x^0}-\sigma^i\frac{\partial\varphi}{\partial x^i}+\sigma^i\frac{\partial\chi}{\partial x^i}&=-i\kappa\varphi-i\kappa\chi,\\ -\frac{\partial\varphi}{\partial x^0}+\frac{\partial\chi}{\partial x^0}+\sigma^i\frac{\partial\varphi}{\partial x^i}+\sigma^i\frac{\partial\chi}{\partial x^i}&=-i\kappa\varphi+i\kappa\chi\end{aligned}\right\} \tag{4.13}$$

and rearranges the terms into:

$$\left.\begin{aligned}\left(\frac{\partial\varphi}{\partial x^0}-\sigma^i\frac{\partial\varphi}{\partial x^i}\right)+\left(\frac{\partial\chi}{\partial x^0}+\sigma^i\frac{\partial\chi}{\partial x^i}\right)&=-i\kappa\varphi-i\kappa\chi,\\ -\left(\frac{\partial\varphi}{\partial x^0}-\sigma^i\frac{\partial\varphi}{\partial x^i}\right)+\left(\frac{\partial\chi}{\partial x^0}+\sigma^i\frac{\partial\chi}{\partial x^i}\right)&=-i\kappa\varphi+i\kappa\chi\end{aligned}\right\} \tag{4.14}$$

then one sees that the differential operators $\nabla$ and $\bar{\nabla}$ operate consistently on $\varphi$ and $\chi$ and one can put the Dirac equation into the equivalent form:

$$\left.\begin{aligned}\bar{\nabla}\varphi &= -i\kappa\chi,\\ \nabla\chi &= -i\kappa\varphi.\end{aligned}\right\} \tag{4.15}$$

This is essentially what we already found for the Weyl representation (4.8) when one sets $\varphi = \psi_d$, $\chi = \psi_u$. Of course, this is because the transformation from the Dirac representation to the Weyl representation amounted to the transformation to an eigenvector basis for the spaces $\Sigma_+$ and $\Sigma_-$. From now on, we shall only refer to the Weyl representation, when necessary, although we now see that one advantage of expressing the Dirac equation in the form (4.15) is that it is no longer necessary to choose a representation for the gamma matrices.

*c. Matrix-valued Dirac wave functions.* Although the wave function $\psi$ for the Dirac electron is usually written as a column matrix that puts $\psi_u$ over $\psi_d$ or $\varphi$ over $\chi$ one can also write it as a 2×2 complex matrix that puts them next to each other as columns:

$$[\psi] = [\varphi \,|\, \chi] = \begin{bmatrix} \varphi^1 & \chi^1 \\ \varphi^2 & \chi^2 \end{bmatrix}. \qquad (4.16)$$

From (4.15), one can then express the Dirac equation in the form:

$$\frac{\partial}{\partial x^0}[\psi] + \sigma^i \frac{\partial}{\partial x^i}[\psi]\sigma^3 = -i\kappa[\psi]\sigma^2. \qquad (4.17)$$

.

*d. Coupling to an external electromagnetic field.* So far, we are only taking about a free Dirac particle. In order to couple the charge and magnetic moment of the electron to an external electromagnetic field, one uses the minimal electromagnetic coupling prescription that takes the form of replacing the momentum operator:

$$p_\mu = i\hbar \frac{\partial}{\partial x^\mu}$$

with

$$p_\mu - (e/c) A_\mu = i\hbar \frac{\partial}{\partial x^\mu} - (e/c) A_\mu,$$

in which $A_\mu = (\phi, A_i)$ represent the components of the electromagnetic potential 1-form $A$, which then defines the electromagnetic field strength 2-form $F$ by its exterior derivative $dA$, which makes:

$$F_{\mu\nu} = \partial_\mu A_\nu - \partial_\nu A_\mu. \qquad (4.18)$$

One recovers the **E** and **B** field from this by imposing a time-space decomposition on Minkowski space, which makes:

$$E_i = F_{0i} = \frac{\partial A_i}{\partial x^0} - \frac{\partial \phi}{\partial x^i}, \qquad \varepsilon_{ijk} B^i = F_{ij} = \partial_i A_j - \partial_j A_i. \qquad (4.19)$$

The Dirac equation then takes the form:

$$[\gamma^\mu(i\hbar\,\partial_\mu - (e/c)\,A_\mu) - mc]\psi = 0. \tag{4.20}$$

or, if we do some rearranging

$$[\gamma^\mu(\partial_\mu + \frac{ie}{\hbar c}A_\mu) + i\kappa]\psi = 0. \tag{4.21}$$

In this form, the coupling of the spin to the external field is by way of the matrix $A_\mu\,\gamma^\mu$, which takes the form:

$$A_\mu\,\gamma^\mu = \begin{bmatrix} 0 & \phi I + A_i\sigma^i \\ \phi I - A_i\sigma^i & 0 \end{bmatrix} \tag{4.22}$$

in the Weyl representation.

Thus, when the matrix is applied to $\psi$, one gets:

$$\begin{bmatrix} \phi\chi + A_i\sigma^i\chi \\ \phi\varphi - A_i\sigma^i\varphi \end{bmatrix}. \tag{4.23}$$

Equations (4.4) and (4.6) then take on the new form:

$$\begin{cases} \left[\dfrac{\partial}{\partial x^0} + \dfrac{ie}{\hbar c}\phi + \sigma^i\left(\dfrac{\partial}{\partial x^i} + \dfrac{ie}{\hbar c}A_i\right)\right]\varphi = -i\kappa\chi, \\ \left[\dfrac{\partial}{\partial x^0} + \dfrac{ie}{\hbar c}\phi - \sigma^i\left(\dfrac{\partial}{\partial x^i} + \dfrac{ie}{\hbar c}A_i\right)\right]\chi = -i\kappa\varphi. \end{cases} \tag{4.24}$$

One must note that the minimal coupling takes a more consistent form in the Dirac representation, although electric and magnetic fields couple to the opposite Pauli spinors. If we introduce the operators:

$$\nabla + \frac{ie}{\hbar c}(\phi + \sigma^i A_i) = \frac{\partial}{\partial x^0} + \frac{ie}{\hbar c}\phi + \sigma^i\left(\frac{\partial}{\partial x^i} + \frac{ie}{\hbar c}A_i\right), \tag{4.25}$$

$$\bar{\nabla} + \frac{ie}{\hbar c}(\phi - \sigma^i A_i) = \frac{\partial}{\partial x^0} + \frac{ie}{\hbar c}\phi - \sigma^i\left(\frac{\partial}{\partial x^i} + \frac{ie}{\hbar c}A_i\right) \tag{4.26}$$

then equations (4.24) take the form

$$\begin{bmatrix} \nabla + \dfrac{ie}{\hbar c}(\phi + \sigma^i A_i) \end{bmatrix}\varphi = -i\kappa\chi, \\ \begin{bmatrix} \bar{\nabla} + \dfrac{ie}{\hbar c}(\phi - \sigma^i A_i) \end{bmatrix}\chi = -i\kappa\varphi, \tag{4.27}$$

which suggests that minimal coupling takes the form:

The relativistic Pauli equation 25$$\nabla \to \nabla + \frac{ie}{\hbar c}(\phi + \sigma^i A_i), \qquad \overline{\nabla} \to \overline{\nabla} + \frac{ie}{\hbar c}(\phi - \sigma^i A_i)$$

in this formulation.

When the up and down components are combined into a matrix, as in (4.16), the system (4.27) takes the form:

$$\left(\frac{\partial}{\partial x^0} + \frac{ie}{\hbar c}\phi\right)[\psi] + \sigma^i\left(\frac{\partial}{\partial x^i} + \frac{ie}{\hbar c}A_i\right)[\psi]\sigma^3 = -i\kappa\psi\,\sigma^1. \qquad (4.28)$$

*e. Charge conjugation.* In order examine the form of the charge conjugation operator $C$, we start with the complex conjugate to equation (4.21):

$$[\gamma^{\mu*}(\partial_\mu - \frac{ie}{\hbar c}A_\mu) - i\kappa]\psi^* = 0. \qquad (4.29)$$

Thus, we have certainly inverted the sign of the charge by this operation. However, we have also replaced the gamma matrices with their complex conjugates and inverted the sign of the rest mass term. Thus, we might look for some transformation $\mathcal{C}$ of $\psi^*$ that restores the equation to the original form with the opposite sign on the charge.

Here, we see how the Majorana representation of the gamma matrices comes in handy, since the fact that they are all imaginary makes $\gamma^{\mu*} = -\gamma^\mu$, and equation (4.29) then becomes:

$$[\gamma^\mu(\partial_\mu - \frac{ie}{\hbar c}A_\mu) + i\kappa]\psi^* = 0. \qquad (4.30)$$

which is the original Dirac equation with the opposite sign on $e$. Thus, the charge conjugate solution $\psi^C$ in this representation is simply the complex conjugate solution $\psi^*$.

More generally, if $\psi^C = \mathcal{C}\psi^*$, so $\psi^* = \mathcal{C}^{-1}\psi^C$, then in order to convert equation (4.29) into the form:

$$[\gamma^\mu(\partial_\mu - \frac{ie}{\hbar c}A_\mu) + i\kappa]\psi^C = 0$$

it is sufficient that:

$$\gamma^{\mu*}\mathcal{C}^{-1} = -\mathcal{C}^{-1}\gamma^\mu. \qquad (4.31)$$

The solution that one uses takes the form:

$$\mathcal{C} = i\gamma^2\gamma^0, \qquad (4.32)$$

so if one defines:

$$\overline{\psi} = \gamma^0\psi^*$$

then charge conjugation takes the form:

$$\psi^C = C\overline{\psi} \qquad (C = i\gamma^2). \qquad (4.33)$$

In the Weyl representation one has:

$$C = \begin{bmatrix} -i\sigma^2 & 0 \\ 0 & i\sigma^2 \end{bmatrix}. \tag{4.34}$$

(Of course, in Majorana representation, one has $C = I$.)

One then sees the reason that the $\sigma^2$ matrices appear so often in the Majorana representation for the gamma matrices, since the transformation from one matrix representation to another is by unitary transformations, and one is looking for a unitary transformation that will essentially take $i\gamma^2\gamma^0$ to the identity matrix.

Since:

$$C\begin{bmatrix} \varphi \\ \chi \end{bmatrix}^* = \begin{bmatrix} -i\sigma^2\varphi^* \\ i\sigma^2\chi^* \end{bmatrix}, \tag{4.35}$$

one can express the effect of charge conjugation on the two types of field as:

$$\varphi^C = -i\sigma^2\varphi^*, \qquad \chi^C = i\sigma^2\chi^*. \tag{4.36}$$

*f. Time-reversal symmetry.* When one replaces $t$ with $-t$ in the Dirac wave function, the corresponding transformation of the field space is (cf., e.g., Bjørken and Drell [**14**]):

$$\psi^T(-t, x^i) = T\,\psi^*(t, x^i). \tag{4.37}$$

In the Weyl representation, one has:

$$T = i\gamma^1\gamma^3 = -\begin{bmatrix} \sigma^2 & 0 \\ 0 & \sigma^2 \end{bmatrix}, \tag{4.38}$$

so:

$$T\begin{bmatrix} \varphi \\ \chi \end{bmatrix}^* = \begin{bmatrix} -\sigma^2\varphi^* \\ -\sigma^2\chi^* \end{bmatrix}, \tag{4.39}$$

and one can say that:

$$\varphi^T = -\sigma^2\varphi^*, \qquad \chi^T = -\sigma^2\chi^*. \tag{4.40}$$

*g. Discrete transformations of matrix wave functions.* When one represents the Dirac bispinor in Weyl form $\psi = [\varphi, \chi]^T$ as a matrix $[\psi] = [\varphi \,|\, \chi]$ the basic transformations that we just discussed can be represented quite simply. One merely looks at how the corresponding transformations of $\psi$ affect the columns of the matrix $[\psi]$.

The parity operator takes $[\varphi, \chi]^T$ to $[\chi, \varphi]^T$, so it should take the matrix $[\varphi \,|\, \chi]$ to $[\chi \,|\, \varphi]$, which makes:

$$[\psi]^P = [\psi]\begin{bmatrix} 0 & 1 \\ 1 & 0 \end{bmatrix} = [\psi]\,\sigma^1. \tag{4.41}$$

The relativistic Pauli equation							27Charge conjugation takes $[\varphi, \chi]^T$ to $[-i\sigma^2\rho^*, i\sigma^2\chi^*]^T$, so it takes $[\psi]$ to $[-i\sigma^2\varphi^* | i\sigma^2\chi^*]$, which makes:

$$[\psi]^C = i\sigma^2[\psi]^* \begin{bmatrix} -1 & 0 \\ 0 & 1 \end{bmatrix} = -i\sigma^2[\psi]^*\sigma^3. \tag{4.42}$$

Time reversal takes $[\varphi, \chi]^T$ to $-[\sigma^2\varphi^*, \sigma^2\chi^*]^T$, so it should take the matrix $[\varphi | \chi]$ to $-[\sigma^2\varphi^* | \sigma^2\chi^*]$, which makes:

$$[\psi]^T = -\sigma^2[\psi]^*. \tag{4.43}$$

This also makes:

$$[\psi]^C = i[\psi]^T\sigma^3. \tag{4.44}$$

Thus, the basic discrete transformations of *P*, *T*, and *C* that are usually discussed for Dirac bispinors are also representable for matrix-valued wave functions, as are Lorentz transformations.

**5. The relativistic Pauli equation.** Although Pauli spinors were first introduced in the context of non-relativistic spinning particles, nevertheless, they can also be treated relativistically. As we saw above, this is because $\mathbb{C}^2$ carries not only a representation (i.e., a linear action) of the non-relativistic group of motions $SU(2)$, it also carries a representation of the relativistic group $SL(2; \mathbb{C})$, which is also by left matrix multiplication. Furthermore, by the use of the Pauli matrices, one can define a representation of real Minkowski space in the complex vector space $M(2; \mathbb{C})$.

*a. The relativistic Pauli equation.* The minimally-coupled Dirac equation, namely:

$$\left[\gamma^\mu\left(\partial_\mu + \frac{ie}{\hbar c}A_\mu\right) + i\kappa\right]\psi = 0, \tag{5.1}$$

can be converted into a second-order equation for $\psi$ by applying the conjugate operator:

$$\gamma^\mu\left(\partial_\mu + \frac{ie}{\hbar c}A_\mu\right) - i\kappa$$

to both sides, which gives:

$$\left[\gamma^\mu\gamma^\nu\left(\partial_\mu + \frac{ie}{\hbar c}A_\mu\right)\left(\partial_\nu + \frac{ie}{\hbar c}A_\nu\right) + \kappa^2\right]\psi = 0. \tag{5.2}$$

In the absence of an external field, this would reduce to the Klein-Gordon equation, but that is entirely due to the symmetry of the second mixed partial derivative operator $\partial_{\mu\nu}$ in $\mu\nu$. This time, we polarize the product of the gamma matrices into:

$$\gamma^\mu \gamma^\nu = \eta^{\mu\nu} + \tfrac{1}{2} \sigma^{\mu\nu}, \tag{5.3}$$

in which:

$$\sigma^{\mu\nu} \equiv [\gamma^\mu, \gamma^\nu] \tag{5.4}$$

are sometimes referred to as the *spin matrices*. In the Weyl representation, they take the form:

$$\sigma^{0i} = 2 \begin{bmatrix} \sigma^i & 0 \\ 0 & -\sigma^i \end{bmatrix}, \quad \sigma^{ij} = -2i\, \varepsilon^{ijk} \begin{bmatrix} \sigma^k & 0 \\ 0 & \sigma^k \end{bmatrix}. \tag{5.5}$$

Thus, the matrices $\sigma^{0i}$ are Hermitian, while the matrices $\sigma^{ij}$ are skew-Hermitian.

We refer to the symmetric part of the first term in (5.2) by the abbreviation:

$$\left(\partial + \frac{ie}{\hbar c} A\right)^2 \equiv \eta^{\mu\nu}\left(\partial_\mu + \frac{ie}{\hbar c} A_\mu\right)\left(\partial_\nu + \frac{ie}{\hbar c} A_\nu\right). \tag{5.6}$$

Note that because this is the product of operators, the left-most operator in the product will differentiate the right-most factor, as well as the wave function that everything is acting on, so one actually has an extra term involving $A_\mu \partial_\nu$ in the final expression:

$$\left(\partial + \frac{ie}{\hbar c} A\right)^2 = \eta^{\mu\nu}\left[\partial_{\mu\nu} + \frac{ie}{\hbar c}(\partial_\mu A_\nu + 2 A_\mu \partial_\nu) + \left(\frac{ie}{\hbar c}\right)^2 A_\mu A_\nu\right] \tag{5.7}$$

The anti-symmetric part of the first term in (5.2) becomes:

$$\tfrac{1}{2}\sigma^{\mu\nu}\left(\partial_\mu + \frac{ie}{\hbar c} A_\mu\right)\left(\partial_\nu + \frac{ie}{\hbar c} A_\nu\right) = \frac{ie}{2\hbar c} \sigma^{\mu\nu}(\partial_\mu A_\nu + A_\mu \partial_\nu + A_\nu \partial_\mu + \frac{ie}{2\hbar c} A_\mu A_\nu)$$

$$= \frac{ie}{2\hbar c} \sigma^{\mu\nu} F_{\mu\nu}.$$

Thus what remains of (5.2) is the equation:

$$\left[\left(\partial + \frac{ie}{\hbar c} A\right)^2 + \frac{ie}{2\hbar c} F_{\mu\nu} \sigma^{\mu\nu} + \kappa^2\right] \psi = 0, \tag{5.8}$$

which one might call the "relativistic Pauli equation" for the Dirac bispinor $\psi$.

When one takes the complex conjugate of this equation, the result is:

$$\left[\left(\partial - \frac{ie}{\hbar c} A\right)^2 - \frac{ie}{2\hbar c} F_{\mu\nu} \sigma^{\mu\nu*} + \kappa^2\right] \psi^* = 0. \tag{5.9}$$

Thus, depending upon the nature of $\sigma^{\mu\nu}$ under complex conjugation, this might also represent the equation for the charge conjugate field. One notes that since $\sigma^{\mu\nu}$ is quadratic in the $\gamma$'s, in the Majorana representation for the $\varphi$'s, it will be real, so, in that case:

$$\left[\left(\partial - \frac{ie}{\hbar c}A\right)^2 - \frac{ie}{2\hbar c}F_{\mu\nu}\sigma^{\mu\nu} + \kappa^2\right]\psi^* = 0, \tag{5.10}$$

which is also the equation for the charge conjugate field, since in that representation $\psi^c = \psi^*$. However, the fact that $\sigma^{\mu\nu}$ is quadratic in the $\gamma$'s also implies that $\sigma^{\mu\nu*}$ commutes with $\mathcal{C}$ in any representation, as well, so one finds that (5.10) is the equation for the charge-conjugate solution in general. Thus, one advantage of the second-order form of the Dirac equation is that is simplifies the nature of the charge conjugation operator.

Our next task will be to express equation (5.8) in terms of a system of equations for a pair of Pauli spinors.

If we wish to examine how the operator in brackets acts on self-dual and anti-self-dual fields then we need to be more specific about the term that couples the external field to the spin of the particle, which takes the form:

$$F_{\mu\nu}\sigma^{\mu\nu} = 2\begin{bmatrix} (E_i - iB_i)\sigma^i & 0 \\ 0 & -(E_i + iB_i)\sigma^i \end{bmatrix} \tag{5.11}$$

in the Weyl representation. In this form, one sees more clearly the difference between the couplings of spin and the electromagnetic field according to the duality type of the wave function.

When the matrix (5.11) is applied to a wave function $\psi = [\varphi, \chi]^T$, the result is:

$$2\begin{bmatrix} (E_i - iB_i)\sigma^i\varphi \\ -(E_i + iB_i)\sigma^i\chi \end{bmatrix}.$$

Thus, equation (5.8) for the Dirac bispinor $\psi$ reduces to the following pair of equations for the Pauli spinors $\varphi$ and $\chi$:

$$\left[\left(\partial + \frac{ie}{\hbar c}A\right)^2 - \frac{ie}{\hbar c}(E_i + iB_i)\sigma^i + \kappa^2\right]\varphi = 0, \tag{5.12}$$

$$\left[\left(\partial + \frac{ie}{\hbar c}A\right)^2 + \frac{ie}{\hbar c}(E_i - iB_i)\sigma^i + \kappa^2\right]\chi = 0, \tag{5.13}$$

which we shall then refer to as the *relativistic Pauli system*.

When one compares these two equations, one finds that the second one is the parity-reverse of the first one, which is consistent with the fact that it is essentially the parity operator that takes the self-dual wave functions to the anti-self-dual ones. In order to see this, one first notes that the parity transformation of the space-time coordinates $x'^0 = x^0$, $x'^i = x^i$ makes:

$$\partial'_0 = \partial_0, \qquad \partial'_i = -\partial_i, \qquad dx'^0 = dx^0, \qquad dx'^i = -dx^i,$$

while the metric, which is quadratic in the spatial 1-forms, remains unchanged.

Thus, one finds that the way the components of the electromagnetic potential 1-form $A = A_0\, dx^0 + A_i\, dx^i$ and field strength 2-form $F = E_i(dt \wedge dx^i) + \tfrac{1}{2} B_{ij}\, dx^i \wedge dx^j$ get altered is to:

$$A = A_0\, dx^0 - A_i\, dx'^i, \qquad F = -E_i(dt \wedge dx'^i) + \tfrac{1}{2} B_{ij}\, dx'^i \wedge dx'^j,$$

which is exactly how the components $E_i$ and $B_i$ have been changed from (5.12) to (5.13).

As for the first term in brackets, one finds that:

$$\begin{aligned}
\eta^{\mu\nu}(\partial'_\mu &+ ie/\hbar c\, A'_\mu)(\partial'_\nu + ie/\hbar c\, A'_\nu) \\
&= (\partial'_0 + ie/\hbar c\, A'_0)^2 - \delta^{ij}(\partial'_i + ie/\hbar c\, A'_i)(\partial'_j + ie/\hbar c\, A'_j) \\
&= (\partial_0 + ie/\hbar c\, A_0)^2 - \delta^{ij}(\partial_i + ie/\hbar c\, A_i)(\partial_j + ie/\hbar c\, A_j), \\
&= \eta^{\mu\nu}(\partial_\mu + ie/\hbar c\, A_\mu)(\partial_\nu + ie/\hbar c\, A_\nu),
\end{aligned}$$

since both terms are quadratic. Thus, the first term is unchanged under parity reversal.

The two equations (5.12) and (5.13) are also related to each other in another simple way: One simply recalls the corresponding equations (4.36) for the charge-conjugate fields, which then makes:

$$\left[\left(\partial - \frac{ie}{\hbar c}A\right)^2 - \frac{ie}{\hbar c}(E_i - iB_i)\sigma^i + \kappa^2\right]\varphi^C = 0, \tag{5.14}$$

$$\left[\left(\partial - \frac{ie}{\hbar c}A\right)^2 + \frac{ie}{\hbar c}(E_i + iB_i)\sigma^i + \kappa^2\right]\chi^C = 0. \tag{5.15}$$

Thus, the charge-conjugated fields satisfy the equations for the opposite duality type with the opposite sign. This is essentially a manifestation of a *CP* symmetry in the system (5.12) and (5.13).

It is important to note that when we go to the pair of second order equations (5.12) and (5.13) we can still include non-zero masses in the equation, whereas this was not the case for the equations for the individual fields $\varphi$ and $\chi$. This is because the first-order differential operators that compose into the second-order operators map the wave functions of one duality type to wave functions of the other, so the composition preserves the duality type, and the reason for restricting to massless waves no longer applies. That is, since the operator $\bar\nabla + ie/\hbar c\,(\phi - A_i\sigma^i)$ maps $\Sigma_+$ to $\Sigma_-$ and the operator $\nabla + ie/\hbar c\,(\phi + A_i\sigma^i)$ maps $\Sigma_-$ to $\Sigma_-$, the two compositions preserve the respective subspaces. This effectively decouples equation (5.8) for the Dirac bispinor $\psi$ into a pair of equations for the Pauli spinors $\varphi$ and $\chi$.

The pair of equations (5.12) and (5.13) for the Pauli spinors $\varphi$ and $\chi$ are, in fact, equivalent to equation (5.8) in terms of Dirac spinor $\psi$, since given a solution $\psi$ to the latter equation one can obtain solutions $\varphi$ and $\chi$ to the system (5.12) and (5.13) by decomposing $\psi$ into its self-dual and anti-self-dual parts, while conversely, given solutions $\varphi$ and $\chi$ to the system (5.12) and (5.13), one can reconstruct a solution $\psi$ to (5.8) by way of:

$$\psi = \begin{bmatrix} \varphi \\ -\varphi \end{bmatrix} + \begin{bmatrix} \chi \\ \chi \end{bmatrix},$$

in the Dirac representation, or $[\varphi, \chi]^T$, in the Weyl representation.

*b. The matrix form of the relativistic Pauli equation.* If one reverts to the representation of a Dirac bispinor in Weyl form $\psi = [\varphi, \chi]^T$ as a 2×2 complex matrix $[\psi] = [\varphi \mid \chi]$, then the equations (5.12) and (5.13) can be consolidated into a single equation for the matrix-valued wave function $[\psi]$:

$$\left[\left(\partial + \frac{ie}{\hbar c} A\right)^2 + \kappa^2\right][\psi] = \frac{ie}{\hbar c} \sigma^i [\psi][F_i], \tag{5.16}$$

in which we have defined the electromagnetic field matrices by:

$$[F_i] = \begin{bmatrix} E_i + iB_i & 0 \\ 0 & -E_i + iB_i \end{bmatrix} = E_i \sigma^3 + iB_i \sigma^0 = \sigma^3(E_i \sigma^0 + iB_i \sigma^3). \tag{5.17}$$

Equation (5.16) then takes the form of a Klein-Gordon type wave equation for $[\psi]$ with a forcing term that amounts to the effect of an external electromagnetic field coupling to the spin of the wave function.

With the definition (4.42), the equation of the charge conjugate field $[\psi]^C$ is, as it should be:

$$\left[\left(\partial - \frac{ie}{\hbar c} A\right)^2 + \kappa^2\right][\psi]^C = -\frac{ie}{\hbar c} \sigma^i [\psi]^C [F_i], \tag{5.18}$$

since $\sigma^2$ anti-commutes with all $\sigma^{i*}$ and:

$$\sigma^3 [F_i]^* = \sigma^3(E_i \sigma^3 - iB_i \sigma^0) = -(E_i \sigma^3 + iB_i \sigma^0)\sigma^3 = -[F_i]\sigma^3.$$

**6. A Lagrangian form for the relativistic Pauli equation.** In order to obtain an action functional for the system of equations (5.12) and (5.13), one must first find one for the system in terms of Dirac bispinors, namely, equation (5.8). The Lagrangian density for that equation takes the form ([1]):

$$\mathcal{L}(\psi, \bar{\psi}, \psi_{,\mu}, \bar{\psi}_{,\mu}) = \tfrac{1}{2}\left\{ \eta^{\mu\nu} \nabla^*_\mu \bar{\psi} \nabla_\nu \psi - \frac{ie}{2\hbar c} F_{\mu\nu} \bar{\psi} \sigma^{\mu\nu} \psi - \kappa^2 \bar{\psi}\psi \right\}, \quad (6.1)$$

in which we are abbreviating the electromagnetically-coupled covariant derivative operator by:

$$\nabla_\mu \equiv \partial_\mu + \frac{ie}{\hbar c} A_\mu \ .$$

Equation (5.8) then becomes the Euler-Lagrange equation that is associated with varying the Lagrangian density (6.1) with respect to $\bar{\psi}$:

$$0 = \frac{\delta \mathcal{L}}{\delta \bar{\psi}} = \frac{\partial \mathcal{L}}{\partial \bar{\psi}} - \partial_\mu \frac{\partial \mathcal{L}}{\partial \bar{\psi}_{,\mu}}, \quad (6.2)$$

which becomes an equation for $\psi$. Conversely, varying $\mathcal{L}$ with respect to $\psi$ then produces the corresponding equation for $\bar{\psi}$.

If one now expresses the Dirac bispinor $\psi$ in the form $\psi = [\varphi, \chi]^T$ and uses the Weyl representation for the spin matrices then after some straightforward, but tedious, computation we find that the Lagrangian density (6.1) takes the somewhat unexpected form:

$$\begin{aligned}
\mathcal{L}(\varphi, \varphi^\dagger, \chi, \chi^\dagger, \varphi_{,\mu}, \varphi^\dagger_{,\mu}, \chi_{,\mu}, \chi^\dagger_{,\mu}) \\
= \tfrac{1}{2}\Big\{ \eta^{\mu\nu} (\nabla^*_\mu \chi^\dagger \nabla_\nu \varphi + \nabla^*_\mu \varphi^\dagger \nabla_\nu \chi) \\
+ \frac{ie}{\hbar c}[(E_i + iB_i)\chi^\dagger \sigma^i \varphi - (E_i - iB_i)\varphi^\dagger \sigma^i \chi] - \kappa^2(\chi^\dagger \varphi + \varphi^\dagger \chi) \Big\},
\end{aligned} \quad (6.3)$$

in which we have discarded an overall factor of 2 that appeared.

Had we chosen the Dirac representation for the spin matrices and a Dirac bispinor of the form $[\varphi + \chi, -\varphi + \chi]^T$ we would have arrived at the same form for the Lagrangian density, although there would have been more steps in the calculations.

One confirms that the system of equations (5.12) and (5.13) comes about from evaluating the pair of equations:

$$\frac{\delta \mathcal{L}}{\delta \chi^\dagger} = 0, \qquad \frac{\delta \mathcal{L}}{\delta \varphi^\dagger} = 0. \quad (6.4)$$

The reason that the form (6.3) of this Lagrangian density should be surprising is that on the surface of things, the system of equations (5.12) and (5.13) appear to be uncoupled in regard to the fields $\varphi$ and $\chi$, while the form of (6.3) seems to involve a bilinear coupling of the fields throughout. In fact, this is solely due to the term $F_{\mu\nu}\sigma^{\mu\nu}$, which couples the external electromagnetic field to the spin of the particle, since in the absence of that term the remaining two equations could be derived from a Lagrangian density that takes the form of the sum of terms that only couple the individual fields to the Hermitian conjugates, namely:

$$\mathcal{L}(\varphi, \varphi^\dagger, \chi, \chi^\dagger, \ldots) = \mathcal{L}(\varphi, \varphi^\dagger, \ldots) + \mathcal{L}(\chi, \chi^\dagger, \ldots)$$

$$= \tfrac{1}{2}\{\eta^{\mu\nu}\nabla^*_\mu\varphi^\dagger\nabla_\nu\varphi - \kappa^2\varphi^\dagger\varphi + \eta^{\mu\nu}\nabla^*_\mu\chi^\dagger\nabla_\nu\chi - \kappa^2\chi^\dagger\chi\}.$$

---

([1]) Although the factor of 1/2 is irrelevant to the equations of motion, nonetheless, it gives the conserved currents a form that is more consistent with one's expectations.

This Lagrangian density would describe only the motion of a charged, but not spinning, particle in an external electromagnetic field, and, in fact, it would suffice to use just $\mathcal{L}(\varphi, \varphi^\dagger)$ or $\mathcal{L}(\chi, \chi^\dagger)$ by themselves, since they both give the same type of wave equation.

Hence, the essential difference between the spinning and non-spinning cases comes down to the nature of the coupling terms:

$$\mathcal{L}_{\text{spin}} = \frac{ie}{\hbar c}[(E_i + iB_i)\chi^\dagger\sigma^i\varphi - (E_i - iB_i)\varphi^\dagger\sigma^i\chi]. \tag{6.5}$$

We can make the Lagrangian (6.3) somewhat more concise by expressing it in terms of the matrix wave function $[\psi]$. One first notes that since $\bar\psi = \psi^\dagger\gamma^0 = [\chi^\dagger, \varphi^\dagger]$ in the Weyl representation, it will correspond to the matrix:

$$[\bar\psi] = \begin{bmatrix} \chi^\dagger \\ \varphi^\dagger \end{bmatrix}. \tag{6.6}$$

One then finds that, for instance, $\bar\psi\psi$ goes to:

$$\text{Tr}([\bar\psi][\psi]),$$

and the original Lagrangian (6.1) now takes the analogous form:

$$\mathcal{L}([\psi], [\bar\psi], \partial_\mu[\psi], \partial_\mu[\bar\psi])$$
$$= \tfrac{1}{2}\text{Tr}\left\{\eta^{\mu\nu}\nabla^*_\mu[\bar\psi]\nabla_\nu[\psi] - \frac{ie}{2\hbar c}[\bar\psi]\sigma^i[\psi][F_i] - \kappa^2[\bar\psi][\psi]\right\}. \tag{6.7}$$

One then obtains equation (5.16) by varying this Lagrangian with respect to $[\bar\psi]$:

$$0 = \frac{\delta\mathcal{L}}{\delta[\bar\psi]} = \frac{\partial\mathcal{L}}{\partial[\bar\psi]} - \partial_\mu\frac{\partial\mathcal{L}}{\partial[\bar\psi]_{,\mu}}. \tag{6.8}$$

**7. Discussion.** We have shown that it is possible to extend the Klein-Gordon equation to account for the coupling of spin with external electromagnetic fields in a manner that is consistent with the Dirac equation, in the sense that any solution of the Dirac equation will, *a fortiori*, be a solution of the relativistic Pauli equation. However, since the relativistic Pauli equation is of second order, while the Dirac equation is of first order, there might be solutions of the former that are not solutions of the latter. Furthermore, one can find the same discrete symmetries that pertain to Dirac bispinors represented in terms of matrix-values wave functions, along with the Lorentz transformations, so one begins to suspect that there is no loss in generality derived from redirecting one's attention from the Dirac equation to the relativistic Pauli equation.

However, since the former equation has long since established itself in the literature of theoretical and experimental particle physics, one must necessarily ask what advantages that one might derive from making that transition, besides a slight reduction in the mathematical abstraction that one derives from eliminating the necessity of introducing gamma matrices and the attendant discussion of the Clifford algebra of real or complex Minkowski space.

It is also our belief that by using wave functions that take their values in $M(2; \mathbb{C})$ one is closer to a direct representation of the physical situation – viz., a charged spinning particle, whereas, since $\mathbb{C}^4$ by itself has no direct physical interpretation, the only way that one can derive a physical interpretation from the Dirac wave function is by a process of "decrypting" the physical observables that it describes the evolution of using the bilinear covariants, which amount to expectation values of the basis elements for $Cl(4; \eta_{\mu\nu})$ for the state that the wave function defines.

We shall address the problem of deriving the equations of motion for the physical observables that are encoded in the wave function of the relativistic Pauli equation in a subsequent paper, since the present one has become voluminous enough simply for the purpose of discussing the details of how the relativistic Pauli equation relates to the Dirac equation, as well as the non-relativistic Pauli equation.